\documentstyle [12pt] {article}

\def\R{{\rm I \!\!\, R}}
\def\be{\begin{equation}}
\def\ee{\end{equation}}
\def\bea{\begin{eqnarray}}
\def\eea{\end{eqnarray}}

\begin{document}

\title{ {\bf  Lie Groups of Conformal Motions acting on Null Orbits}}
\author{ A.M. Sintes\thanks{Max-Planck-Institut f\"ur Gravitationsphysik.
Albert-Einstein-Institut. D-14473 Potsdam. Germany.},
 A.A. Coley\thanks{ Department of Mathematics, Statistics and
 Computing Science,  Dalhousie University, Halifax, NS. Canada B3H
3J5.} and  J. Carot\thanks{Departament de F\'{\i}sica.
 Universitat de les Illes Balears.
E-07071 Palma de Mallorca. Spain. }}
\date{}
 \maketitle

\begin{abstract}

Space-times admitting a 3-dimensional Lie group of conformal motions $C_3$
acting on null orbits are studied. Coordinate expressions for the metric and
the conformal Killing vectors (CKV) are  provided (irrespectively of the
matter content) and then all possible perfect fluid solutions are found, 
although
none of these verify the weak and dominant energy conditions over the whole
space-time manifold. \end{abstract}

In this letter we shall consider
space-times $(M,g)$
 admitting a maximal  three-parameter conformal group $C_3$ containing
 an Abelian
two-parameter subgroup of isometries $G_2$ whose orbits $S_2$ are spacelike,
diffeomorphic to $\R^2$ and admit orthogonal two-surfaces.
Furthermore, we shall assume that the $C_3$  acts transitively on null
orbits $N_3$, thus complementing a previous paper \cite{Ali3} in which the case
of null conformal orbits was explicitly excluded.
In particular, in
 this letter we shall provide the coordinate expressions for
 the metric and the CKV for each Lie algebra structure and give
{\it all} possible perfect fluid solutions.

A few remarks concerning Lie groups acting on null orbits are in order here.
In most cases  the study of null orbits has been restricted to  isometries
only. It is a well known fact that   groups $G_r$, $r\ge 4$, 
acting on $N_3$
 have at least one subgroup $G_3$
 acting on $N_3$, $N_2$ or $S_2$ \cite{Kramer}. 
 In the case in which the subgroup
  $G_3$ 
acts on $S_2$,
the space-time is a
 LRS model, and the $G_r$
 admits either a different subgroup $G_3$ acting 
on $N_3$ or a null Killing vector (KV) \cite{Barnes73}. The case $G_3$ 
acting on $N_2$
was studied by Barnes \cite{Barnes79}; the group $G_3$ is then of Bianchi type
$II$ and perfect fluid solutions are excluded since the metric 
leads to a Ricci tensor whose Segre type is not that of a perfect fluid.
Another case that has been considered in the literature is that
of a $G_3$ acting 
on $N_3$ in which $R_{ab}k^ak^b=0$,
and this condition excludes perfect fluid sources with $\mu+p\not=0$.
It is also known that perfect
fluid solutions cannot admit a non-twisting $(w=0)$ null KV
except when $\mu+p=0$. The algebraically special perfect fluid solutions with
a twisting null KV are treated by Wainwright \cite{Wain70}, and
 they
admit an Abelian group $G_2$.
Space-times admitting a null CKV have been studied recently by Tupper
 \cite{tup}. He has found that, for perfect fluid and null radiation,
 non-conformally flat  space-times admitting a null CKV are algebraically
special; furthermore, if one assumes the CKV to be proper (non-homothetic)
then the only possibilities are those solutions in which the line element
admits a multiply transitive group of isometries $G_3$ acting on two-spaces
of constant curvature.

One might get the impression that space-times
admitting  a three-dimensional Lie group of conformal motions $C_3$ acting
on null orbits (i.e., the case under consideration here)
might not admit any 
 perfect fluid solutions, since the line element of these space-times
is, by the theorem of Defrise-Carter \cite{Defrise}, conformally related to one
admitting a $G_3$ acting on null orbits and 
such space-times,
 as we have pointed out
above, do not admit perfect fluid solutions. 
However,  we will show that this is
not the case. Indeed, a conformal scaling   changes the
algebraic structure of the Ricci
tensor.
Nevertheless, we find that there are only a few perfect fluid solutions
possible.

The classification of all possible Lie algebra structures for  ${\cal C}_3$
 was given in \cite{Ali3} where
coordinates were adapted so that the line element  associated with the metric 
$g$
can be written as
\be
ds^2=e^{2F}\{-dt^2+dr^2+Q[H^{-1}(dy+Wdz)^2+Hdz^2]\}\ ,
\label{n6}
\ee
where $F$, $Q$, $H$ and $W$ are all functions of $t$ and $r$ alone.
(The precise hypotheses leading to this classification were given 
explicitly in Ref. \cite{Ali3}.)

If the conformal 
algebra ${\cal C}_3$ belongs to the family A 
(i.e., the commutator between the CKV and each  KV is a KV),
 it was shown in \cite{Ali3}
 that, for null conformal orbits, one can always bring the CKV, $X$,
 to the form 
\be
 X=\partial_t+\partial_r+X^y(y,z)\partial_y+X^z(y,z)\partial_z \ , \label{n2}
\ee
where $X^y(y,z)$ and $X^z(y,z)$ are linear functions of their arguments to be
determined from the commutation relations between $X$ and the KVs.
Considering now the conformal Killing equations
for the CKV (\ref{n2}) and the metric (\ref{n6}),  
 for each possible group type,
 one obtains the following forms for $X$ and the metric
functions $F$, $Q$, $H$, and $W$ appearing in (\ref{n6}) as follows:
\bea
(I)&\,& Q=q(t-r), \quad H=h(t-r), \quad W= w(t-r), \nonumber\\
&\,& X=\partial_t+\partial_r. \\
(II)&\,& Q=q(t-r), \quad H=h(t-r), \quad W= w(t-r)-{t+r \over 2}, \nonumber\\
&\,& X=\partial_t+\partial_r+z\partial_y.\\
(III)&\,& Q=e^{-{t+r \over 2}}q(t-r), \quad H=e^{{t+r \over 2}} h(t-r), \quad
W=e^{{t+r \over 2}} w(t-r), \nonumber\\ &\,& X=\partial_t+\partial_r
+y\partial_y. \label{n13}\\
 (IV)&\,& Q=e^{-(t+r)}q(t-r), \quad H=h(t-r), \quad W=
w(t-r)-{t+r \over 2}, \nonumber\\ &\,& X=\partial_t+\partial_r
+(y+z)\partial_y+z\partial_z.\\
 (V)&\,& Q=e^{-(t+r)}q(t-r), \quad
H=h(t-r), \quad W= w(t-r), \nonumber\\ &\,&
X=\partial_t+\partial_r +y\partial_y+z\partial_z. \\ 
(VI)&\,&
Q=e^{-(1+p){t+r \over 2}}q(t-r), \quad H=e^{(1-p){t+r \over 2}}h(t-r),  \quad
W=e^{(1-p){t+r \over 2}} w(t-r), \nonumber\\ &\,& X=\partial_t+\partial_r
+y\partial_y+pz\partial_z\quad (p\not=0,1).\label{n16} \\ 
(VII)&\,& Q=e^{-p{t+r \over 2}}q(t-r),\quad 
c=c(t-r), \quad g=g(t-r), \nonumber\\
 &\,& H={{\sqrt{4-p^2}\over 2}\over \sqrt{1+c^2+g^2} 
+c\cos(\sqrt{4-p^2}{t+r \over 2}) +g\sin(\sqrt{4-p^2}{t+r \over 2})} , 
\nonumber\\ &\,& W={p\over 2}+{{\sqrt{4-p^2}\over 2}
[c\sin(\sqrt{4-p^2}{t+r \over 2})-g\cos(\sqrt{4-p^2}{t+r \over 2})] \over
\sqrt{1+c^2+g^2} +c\cos(\sqrt{4-p^2}{t+r \over 2}) +g\sin(\sqrt{4-p^2} {t+r
\over 2})}, \nonumber\\ &\,& X=\partial_t+\partial_r
-z\partial_y+(y+pz)\partial_z\quad (p^2<4). \label{n18}\eea
In all of these cases $F=F(t,r)$ and the conformal factor $\Psi$ is given by
\be
\Psi=F_{,t}+F_{,r} \   .
\ee

Note that these results are completely independent of
the Einstein field equations and therefore of
the assumed energy-momentum tensor. 
Furthermore, it is easy to prove that family B
(i.e., the case in which the commutator between the CKV and 
at least one KV is a proper CKV)
 cannot admit
CKV acting on null orbits (the proof can be found  in
\cite{tesis}).

Let us now study possible
 perfect fluid solutions. For a maximal $C_3$, with
a proper CKV, all possible solutions have been found.
We will summarize the results obtained for the different metrics 
(the details can be obtained from Ref. \cite{tesis}).
For type $I$ (i.e., the case in which $X$ is a null CKV), we find that the
space-time always admits a further KV tangent to the Killing orbits, and the
metric then admits a multiply transitive group $G_3$ of isometries. This 
result is consistent with Tupper's analysis \cite{tup}. 
For types $II$ and $IV$, either $X$ is not a proper CKV or it does not 
correspond to a perfect fluid solution (i.e., wrong Segre type). For 
types $V$ and $VII$ it can be shown that either $C_3$ is not maximal 
or $X$ is not a proper CKV
 (see \cite{tesis} for
details). 
Therefore, perfect fluid solutions under the previous hypotheses 
can only occur for  the types $III$ and $VI$.

{\bf Type $VI$} ( including  type $III$ for $p=0$): 

We make the coordinate transformation  $u=t+r$ and $v=t-r$, so that we have
$h=h(v)$ and $q=q(v)$. The field equations yield
\be
W=0 \ ,
\ee
\be
F=f(x)+{1\over 2}{1+p \over 1-p}\ln h -{1\over 2} \ln q \ ,\quad x\equiv
u-{2\over 1-p}\ln h \ , \label{n60}
\ee
\be
0=\left\{ {q_{,v}h_{,v} \over qh}+{h_{,vv}\over h}\right\}\Sigma_0
+\left({h_{,v}\over h}\right)^2\Sigma_1\ ,\label{n61}
\ee
where
\bea
\Sigma_0&\equiv&-1+p^4+4f_{,x}-4pf_{,x}+4p^2f_{,x}-4p^3f_{,x}+8f_{,x}^2
-8p^2f_{,x}^2 \nonumber\\
 & -&32f_{,x}^3+32pf_{,x}^3-8f_{,xx}+8p^2f_{,xx}+32f_{,xx}f_{,x}
-32pf_{,xx}f_{,x} \ ,
\eea
\bea
\Sigma_1&\equiv &2+2p+2p^2+2p^3-16f_{,x}-8pf_{,x}-16p^2f_{,x}-8p^3f_{,x}
+32f_{,x}^2 +16pf_{,x}^2 \nonumber\\
&+&48p^2f_{,x}^2-64pf_{,x}^3-16f_{,xx}+16pf_{,xx}-32pf_{,xx}+64pf_{,xx}f_{,x}\ ,
\eea
 and $h_{,v}=0$ is excluded since the solution
 does not correspond to a perfect fluid. Therefore, two possibilities arise:
\bea
&{\rm i})& \quad \Sigma_0=0, \quad \Sigma_1=0 \ , \nonumber \\
&{\rm ii})&\quad {\displaystyle q_{,v}h_{,v} \over\displaystyle qh}+
{\displaystyle h_{,vv}\over\displaystyle h}=a \left({\displaystyle
h_{,v}\over\displaystyle h}\right)^2 \qquad (a={\rm const})\ . \nonumber 
\eea
 In the first case $f_{,x}$ must be a constant, and
therefore the CKV is not proper. In the second case we have that
\be
{q_{,v}\over q}=a{h_{,v}\over h}-{h_{,vv}\over h_{,v}}\ , \label{n64}
\ee
which can be integrated to give
\be
q={h^a \over h_{,v}}\ , \label{n65}
\ee
and equation (\ref{n61}) reduces to:
\be
1={f_{,xx}[f_{,x}32(ap-a-2p)+8(2-p^2a-2p+4p^2+a)]\over [4f_{,x}-p-1]
[f_{,x}^2 8(ap-a-2p)+f_{,x}8(p^2+1)+a-ap+ap^2-ap^3-2-2p^2]}\  .\label{n66} \ee
It is convenient to 
further divide the analysis into three sub-cases. \hfill\break

\underline{Sub-case (a):} $a=2p/(p-1)$.
Equation (\ref{n66})  can be readily integrated to give
\be
f={p+1 \over 4}x- {(1-p)^2\over p^2+1}{1\over 2}\ln\vert x\vert +c\ , \quad
c={\rm const} \ . \label{n69}
\ee
We notice that for $p=-1$ there exists a third KV of the form
\be
\zeta=\left( {1\over 2}+{1\over 2}{h\over h_{,v}}\right) \partial_t+
\left({1\over 2}-{1\over 2}{h\over
h_{,v}}\right) \partial_r+y\partial_y-z\partial_z \ .\label{n70}
\ee
 \hfill\break

\underline{Sub-case (b):} $a=2/(1-p$).
When $p=-1$ the solution is a particular case of  sub-case (a). The
remaining cases may  now be integrated giving:
\be
f=-\ln\vert 1-e^{-(1+p)x/4} \vert +c \ ,\quad
c={\rm const} \ .\label{n73}
\ee
We note that in this sub-case there exists a further KV
\be
\zeta=\left( {1\over 2}+{1-p\over 4}{h\over h_{,v}}\right)\partial_t+ \left(
{1\over 2}-{1-p\over 4}{h\over h_{,v}}\right)\partial_r+ {1-p\over
2}y\partial_y -{1-p\over
2}z\partial_z \ , \label{n74}
\ee
which violates our requirement of a maximal three-dimensional
conformal  group $C_3$.
 \hfill\break

\underline{Sub-case (c)}:  finally we consider the possibility
$a\not=2p/(p-1)$ and $a\not=2/(1-p)$.
The solution of (\ref{n66}) is then given implicitly by
\be
x=\gamma_1\ln\vert f_{,x}-\beta_0\vert + \gamma_2\ln\vert f_{,x}-\beta_+\vert +
\gamma_3\ln\vert f_{,x}-\beta_-\vert \ , \label{n79}
\ee
where
\be
\beta_0={p+1\over 4}\ ,\quad
\beta_{\pm}={-2(p^2+1)\pm\sqrt{2(p^2+1)(1-p)^2(a^2-2a+2)}\over 4(ap-a-2p)}\ ,
\label{n76}
\ee
and $\gamma_i$, $i=1,2,3$,  are constants satisfying 
$\gamma_1+\gamma_2+\gamma_3=0$.

A careful analysis of the energy conditions shows that for all cases (i.e.,
for all values of the parameters
$a$ and $p$) the solutions can only satisfy the energy
conditions over certain open domains of the manifold (see \cite{tesis}).


\begin{thebibliography}{99}
\bibitem{Ali3} J. Carot, A.A. Coley and A.M. Sintes,  Gen. Rel. Grav., {\bf
28}(1996)311.
\bibitem{Kramer} D. Kramer, H. Stephani, M.A.H. MacCallum and E. Herlt, {\em
Exact Solutions of Einstein's Field Equations}, Deutscher Verlag der
Wissenschaften, Berlin (1980).
\bibitem{Barnes73} A. Barnes,  Commun. Math. Phys., {\bf 33}(1973)75.
\bibitem{Barnes79} A. Barnes,  J. Phys. A, {\bf 12}(1979)1493.
 \bibitem{Wain70} J. Wainwright,  Commun.
Math. Phys., {\bf 17}(1970)42.  
\bibitem{tup} B.O.J. Tupper, private communication (1997).
\bibitem{Defrise} L. Defrise-Carter, Commun. Math. Phys., {\bf
40}(1975)273.
\bibitem{tesis} A.M Sintes, Ph.D. Thesis, Universitat de les Illes Balears
(1996).
 \end{thebibliography}
\end{document}